\documentclass[twocolumn,showpacs,preprintnumbers,amsmath,amssymb,prl]{revtex4}

\usepackage{graphicx}% Include figure files
\usepackage{dcolumn}% Align table columns on decimal point
\usepackage{bm}% bold math

\newcommand{\abs}[1]{\ensuremath{\left|#1\right|}}

\newcommand{\ket}[1]{\ensuremath{\left|\left. #1\right>\right.}}
\newcommand{\bra}[1]{\ensuremath{\left<\left. #1\right|\right.}}
\newcommand{\braket}[2]{\ensuremath{\left<\left.#1\left| #2\right>\right.\right.}}
\newcommand{\ketbra}[2]{\ensuremath{\ket{#1}\!\!\!\bra{#2}}}

\newtheorem{lemma}{Lemma}

\begin{document}

%\preprint{}

%\title{How the collapse of the wavefunction occurs without postulation or approximation}
\title{Exactly decohering quantum measurement without environment}

\author{Eric A. Galapon}
\email{eagalapon@up.edu.ph}
\affiliation{Theoretical Physics Group, National Institute of Physics, University of the Philippines, Diliman Quezon City, 1101 Philippines}

\date{\today}

\begin{abstract}
Current quantum orthodoxy claims that the statistical collapse of the wave-function arises from the interaction of the measuring instrument with its environment through the phenomenon known as environment induced decoherence. Here it is shown that there exists a measurement scheme that is exactly decohering without the aid of an environment. The scheme relies on the assumption that the meter is decomposable into probe and pointer, with the probe taken to be inaccessible for observation. Under the assumption that the probe and the pointer initial states are momentum limited, it is shown that coherences die out within a finite interval of time and the pointer states are exactly orthogonal. These lead to the fundamental realization that dispersion of correlation does not require an external infinite number of degrees of freedom. An internal one degree of freedom is already sufficient to delocalize the correlations and leave its subparts in a classical mixed state, so that decoherence may occur even for isolated measuring instruments. 
\end{abstract}


\maketitle

Our everyday experience with measuring instruments registering unambiguous read-outs runs conflict with the unitary dynamics of quantum mechanics. The quantum equation of motion due to the unitary Schr\"{o}dinger equation evolves an initial definite meter reading (say zero) to a linear superposition of all possible readings that are in correlation with the possible values of the property of the system being measured; this leaves the meter in a state without a definite reading, contrary to our experience. This comes about because the interaction of the meter and the system during measurement entangles them into an inseparable unit in a definite state--with neither the meter nor the system having a separate state of its own. Within the confines of standard quantum mechanics, the emergence of definite outcome from this entangled state is addressed by means of a process known as quantum decoherence. The environment induced decoherence theory (EIDT) provides a mechanism for decoherence to occur \cite{zeh1970,zeh1973,zurek1981,zurek1982,zurek2003,zurek2009,schlosshauer,schlosshauer2}. It is anchored on the realization that the meter is not isolated from the environment; and it is the interaction between the environment and the meter that enforces decoherence. However, EIDT is being criticized for not being ``a well defined process'' for the following reasons: (i) Suppression of coherences is approximate only; (ii) the induced pointer basis is only approximately orthogonal; (iii) there is arbitrariness in the boundary between the meter and environment \cite{pessoa,ballentine,tanona,wallace}.  

Here I show that decoherence can be achieved without the environment, relying only on the assumption that the meter has to be decomposed into probe and pointer, with the probe taken to be inaccessible for observation. While current models of EIDT requires infinite time for quantum decoherence to occur,
here exact decoherence occurs at a finite time. Moreover, while EIDT leads to only approximate orthogonal pointer states, here pointer states are exactly orthogonal.  These lead to the fundamental realization that dispersion of correlation does not require an external infinite number of degrees of freedom. An internal one degree of freedom is already sufficient to delocalize the correlations and leave its subparts in a classical mixed state, so that decoherence may occur even for isolated measuring instruments.

In general the state of a quantum system is described in terms of the density matrix to accommodate both pure and mixed states. Pure states satisfy the superposition principle, and they are characterized by the existence of a measurement whose outcome is predictable with certainty. On the other hand, mixed states are weighted superpositions of pure states, and they are characterized by the absence of measurement whose outcome is predictable with certainty. Quantum interference of mutually exclusive outcomes are manifested by the off-diagonal elements---the coherences---of the density matrix. Full coherence is the hallmark of pure states, and degraded coherence is that of mixed states. An extreme class of states constitutes those that lack coherences at all, i.e. diagonal mixed states. For such states probabilities computed from them satisfy the classical rules of probability, and hence may be consistently interpreted as classical states admitting ignorance interpretation. Now the eventual disappearance of the coherences of a pure state while leaving the diagonal of the corresponding density matrix intact is referred to as decoherence. Any decoherence process then maps a pure quantum state into a mixed classical state. Such a process cannot be unitary, because unitary dynamics maps a pure state into another pure state. 

However, a non-unitary evolution can be achieved from the unitary Schr\"{o}dinger equation by entangling the system with another system through a unitary interaction, and then ignoring the auxiliary system.  Entanglement allows information to leak out of the system and ignoring the auxiliary system leaves the system in a mixed state. In environment induced decoherence theory the measuring instrument is immersed in a environment, which is, by definition, a quantum system with an arbitrarily large or infinite number of unobservable degrees of freedom. Decoherence is accomplished by tracing out the environment, leaving the system and the measuring instrument in a mixed state. In this paper our approach is to endow the measuring instrument an inaccessible internal degree of freedom which does not play a direct role in the registration of measurement outcome, but nevertheless is capable of dispersing quantum correlations. 

Here we will consider the measurement of a non-degenerate observable $A$ of a finite dimensional quantum system $\mathcal{S}$. To set the stage for the development to follow, we briefly review the quantum measurement scheme due to von Nuemann for such an observable. In the von Neumann scheme, the system $\mathcal{S}$ is coupled to a meter $\mathcal{M}$. The system and the meter are both treated as quantum objects with respective Hilbert spaces $\mathcal{H}_S$ and $\mathcal{H}_M$, with the system plus meter described by the tensor product Hilbert space $\mathcal{H}_S\otimes\mathcal{H}_M$. The system and the meter are assumed to be initially prepared in the uncorrelated pure states, $\ket{\psi_0}$ and $\ket{\phi_0}$, respectively, so that the composite initial state is given by $\ket{\Psi}=\ket{\psi_0}\!\!\otimes\!\ket{\phi_0}$. The measurement is carried out by coupling the system observable $A$ with a meter observable $B$, and is modeled by the measurement Hamiltonian $H_{M,\mbox{von}}=\lambda A\otimes B$, where $\lambda$ is a coupling constant (see Methods) \cite{busch,von}. Then the state of the system and meter after the measurement is $\ket{\Psi_f}=e^{ i \lambda A\otimes B/\hbar}\ket{\psi_0}\!\!\otimes\!\ket{\phi_0}$. 

Let $\ket{\varphi_k}$ be the eigenvectors of $A$ and $a_k$ the corresponding eigenvalues. Then $e^{i \lambda A\otimes B/\hbar} = \sum_{k} \ket{\varphi_k}\!\!\!\bra{\varphi_k} \otimes e^{i \lambda a_k B/\hbar}$, from which we obtain the final state after the measurement
\begin{equation}
\ket{\Psi_f}=\sum_{k} \braket{\varphi_k}{\psi_0} \ket{\varphi_k} \otimes \ket{\phi_k},
\end{equation}
where $\ket{\phi_k}=e^{i \lambda a_k B/\hbar} \ket{\phi_0}$. This state is called the pre-measurement state. The density matrix for the final state is $\rho_f=\ketbra{\Psi_f}{\Psi_f}$
\begin{equation}\label{che}
\rho_f = \sum_{k,l} \braket{\varphi_k}{\psi_0} \braket{\psi_0}{\varphi_l} \ketbra{\varphi_k}{\varphi_l} \otimes \ketbra{\phi_k}{\phi_l}.
\end{equation}
Von Nuemmann postulated that aside from the unitary evolution there is a non-unitary reduction of the state to an appropriate mixture,
\begin{equation}\label{cheche}
\rho_f\mapsto \rho_f'=\sum_{k} |\braket{\varphi_k}{\psi_0}|^2 \ketbra{\varphi_k}{\varphi_k} \otimes \ketbra{\phi_k}{\phi_l}
\end{equation}
This reduction of the density matrix from an initial pure state to a mixed state is what will refer to as the statistical collapse of the wave-function in this paper. The need for the reduction of the density matrix for a completed measurement is discussed in \cite{zurek1981,zurek1982}. 

We now devote the rest of the paper to show that there exists a universal measurement scheme that solves the statistical collapse of the wave-function without postulation. The key to our treatment is the realization that the measuring instrument has to be decomposed in two parts: the probe and the pointer. In orthodox quantum mechanics, these two parts are lumped into one system. The entire Hilbert space comprising the measurement scheme is then $\mathcal{H}_S\otimes\mathcal{H}_{Pr}\otimes\mathcal{H}_{Po}$, where $\mathcal{H}_S$ is the system Hilbert space, $\mathcal{H}_{Pr}$ is the probe Hilbert space, and $\mathcal{H}_{Po}$ is the pointer Hilbert space. We will find it necessary that the probe and pointer Hilbert spaces are infinite dimensional. The constraints on the infinite dimensionalities of the probe and pointer Hilbert spaces will be demonstrated as necessary to achieve exact reduction and existence of a projection valued measure pointer observable. Incidentally the infinite dimensionality of the pointer Hilbert space will accommodate any dimensionality of the system, so that the measuring instrument works for all systems.

We assume a measurement model similar to that of von-Neumann's scheme with the measurement Hamiltonian 
\begin{equation}
H_M=\left[\alpha A\otimes Q \otimes \mathbb{I}_{Po} + \beta \mathbb{I}_{S}\otimes P \otimes B\right],
\end{equation}
where $Q$ and $P$ are the respective position and momentum operators of the probe, and $\alpha$ and $\beta$ are real coupling constants which we take to be arbitrary at the moment; $A$ is the observable to be measured and $B$ is a pointer observable. The necessity of the infinite dimensionality of the probe Hilbert space is made clear by the introduction of the position $Q$ and momentum $P$ operators in the Hamiltonian. 
Notice that there is no direct coupling between the system and the pointer in the Hamiltonian. Their interaction is mediated through the probe. We take the initial state of the combined system to be the pure product state $\ket{\Psi}=\ket{\psi_0}\!\otimes\!\ket{\Psi_{P}}\!\otimes\ket{\Phi_0}$, where $\ket{\psi_0}$, $\ket{\Psi_{P}}$, $\ket{\Phi_0}$ are the initial states of the system, probe and pointer, respectively. Then the state after the measurement is
\begin{equation}
\ket{\Psi_f}=U\ket{\psi_0}\!\otimes\!\ket{\Psi_{P}}\!\otimes\ket{\Phi_0}
\end{equation}
where the unitary time evolution operator $U$ is given by
\begin{equation}\label{evo}
U=\exp\left(-\frac{i}{\hbar}\alpha A\otimes Q \otimes \mathbb{I}_{Po} - \frac{i}{\hbar} \beta \mathbb{I}_S \otimes P \otimes B\right) .
\end{equation}

To unravel the physical content of the final state, the evolution operator (\ref{evo}) is factored in the form 
\begin{equation}
U=e^{\frac{i}{2\hbar}\alpha\beta A\otimes \mathbb{I}_{Pr}\otimes B} \cdot e^{-\frac{i}{\hbar} \alpha A\otimes Q \otimes \mathbb{I}_{Po}}\cdot e^{-\frac{i}{\hbar}\beta \mathbb{I}_S\otimes P \otimes B},
\end{equation}
where we have used the commutation relation $[A\otimes Q \otimes \mathbb{I}_{Po},\mathbb{I}_S\otimes P \otimes B]=i\hbar A\otimes \mathbb{I}_{Pr}\otimes B$ to obtain this result. With the obvious correspondences, we write $U=U_{AB}U_{AQ}U_{PB}$. In this form we find that the total evolution is equivalent to a compound action, with the system-pointer interaction explicitly emerging. We now chose $\beta$ in such a way that the induced coupling of $A$ and $B$ in $U_{AB}$ is independent of the coupling of $A$ and $B$ to the probe, in particular, $\beta=2\lambda/\alpha$, where $\lambda>0$ is now the coupling constant between $A$ and $B$, and is independent of $\alpha$.  

Since the probe is just there to mediate the interaction between the system and the pointer and since we do not observe it, we trace it out to obtain the reduced density matrix for the combined system and pointer,
\begin{equation}\label{ke}
\rho_{S\otimes M}= U_{AB} \rho_0^*  U_{AB}^{\dagger},
\end{equation}
where $U_{AB}$ now is simply given by $U_{AB}=e^{i \alpha\beta A\otimes B/\hbar}$ and
\begin{equation}\label{keke}
\rho_0^*= \mbox{Tr}_{Pr}\left[U_{AQ}U_{PB}\ketbra{\Psi_f}{\Psi_f}U_{PB}^{\dagger} U_{AQ}^{\dagger}\right] U_{AB}^{\dagger},
\end{equation}
Equation is equivalent to a quantum operation $\epsilon$ acting on the initial state $\rho_{S\otimes Po} =\ketbra{\Psi_0}{\Psi_0}\otimes \ketbra{\Phi_0}{\Phi_0}$ of the system and the pointer, i.e. $\rho_0^*=\epsilon(\rho_{S\otimes Po})$. 

Expression (\ref{keke}) holds for all pointer observables, but we now require $B$ to posses a simple spectrum spanning the real line. This requirement implies that $B$ is necessarily unbounded and that the Hilbert space of the pointer is necessarily infinite dimensional. Let $\ket{q}$ and $\ket{b}$ be the generalized eigenvectors of $Q$ and $B$, respectively. Then 
\begin{eqnarray}
\rho_0^* &=& \sum_{k,j} \braket{\varphi_k}{\Psi_0} \braket{\Psi_0}{\varphi_l} \,\ketbra{\varphi_k}{\varphi_l}\nonumber \\
& & \hspace{-4mm} \otimes \int\!\!\!\!\int \mbox{d}b \mbox{d}b' \, I_{k,l}(b,b') \braket{b}{\Phi_0} \braket{\Phi_0}{\Phi_0} \ketbra{b}{b'}
\end{eqnarray}
where
\begin{eqnarray}
I_{kl}(b,b')&=&\int dq\, e^{\frac{i}{\hbar} \alpha q (a_k-a_l)} \nonumber \\
&& \hspace{-4mm}\times \braket{q-2\lambda b/\alpha}{\Psi_{Pr}} \braket{\Psi_{Pr}}{q-2\lambda b'/\alpha}.
\end{eqnarray}
The function $I_{kl}(b,b')$ controls the coherences, their presence or absence. 

At this point we can now make observations on the behavior of $I_{kl}$. Coherences are suppressed if they vanish for $k\neq l$. The expression is a Fourier transform in the coupling parameter $\alpha$, and it is a well-known theorem that the Fourier transform vanishes as $\alpha\rightarrow\infty$, i.e. $\lim_{\alpha\rightarrow\infty} I_{k,l}=0$. This provides a mechanism for the disappearance of the coherences. In the limit
\begin{equation}
\lim_{\alpha\rightarrow\infty} \rho_{S\otimes M} = \sum_{k} |\braket{\varphi_k}{\psi_0}|^2 \ketbra{\varphi_k}{\varphi_k} \otimes \ketbra{\phi_k}{\phi_k},
\end{equation}
which is the desired reduced density matrix. Then von Neumann reduction of the density matrix arises in the limit of infinite coupling between the probe and the system. However, while we have obtained the desired statistical collapse of the wave-function, it is not what we are after for. The reason is that the collapse occurs only in the unphysical limit of infinite coupling between the system and the probe, and zero coupling between the probe and the pointer. 

We may address this by imposing $\alpha$ to be arbitrarily large but finite, and argue that the coherences are small. For example, we may prepare the probe to be initially in a harmonic oscillator ground state, $\braket{q}{\Psi_{Pr}}=(\mu\omega/\hbar\pi)^{1/4} e^{-\mu \omega q^2/2\hbar}$. For this case, we have $I_{kl}=e^{-\alpha^2 (a_k-a_l)^2/4\mu\omega\hbar} f(b,b')$, where $f(b,b')$ is independent of $\alpha$; we obtain Gaussian suppression of the coherences. For such suppression, we can then argue that for all practical purposes the coherences are zero. But this suffers from the same criticism as that on the environment induced decoherence that the loss of coherence is only approximate.

We demand the vanishing of the coherences at finite coupling interaction $\alpha$. We now show that under a specific set of initial states of the probe the coherences vanish at finite coupling strength. To show we exploit the fact that $I_{kl}$ is a Fourier transform. Demanding that $I_{kl}$ vanish at finite $\alpha$, we are demanding that $I_{kl}$ has a compact support: It is non-vanishing only in a finite interval of $\alpha$. The fact that $I_{kl}$ is a Fourier transform, combined with the well-known Paley-Wiener theorem involving exponential type functions, the function $I_{kl}$ can be made to vanish for some finite $\alpha$ if the probe wave function $\braket{q}{\Psi_{Pr}}$ has extension in the entire complex plane, $\braket{z}{\Psi_{Pr}}$, which is an entire function of exponential type $\kappa_0>0$. 

A complex function $f(z)$ is entire if its series expansion has infinite radius of convergence; and $f(z)$ is exponential of type $\tau$ if, for sufficiently large $|z|$, we have the asymptotic inequality $f(z)<e^{\tau |z|}$ (\cite{boas}). A wave-function $\psi(x)$ in $L^2(\mathbb{R})$ is momentum limited if $\psi(x)=\int_{-\kappa_0}^{\kappa_0} \mbox{e}^{i x k} \tilde{\psi}(k)\mbox{d}k$ for some $0<\kappa_0<\infty$. It is known that such a function has an extension $\psi(z)$ in the complex plane that is both entire and exponential of type $\kappa_0$ \cite{boas}. Now central to the proof of the vanishing of the coherences at finite coupling strength is the  
\begin{lemma}
	Let $f(z)$ be entire and exponential of type $\tau>0$, and $\int_{-\infty}^{\infty} |f(x)|\, \mbox{d}x=M<\infty$; then $\int_{-\infty}^{\infty} e^{i a x} f(x)\, \mbox{d}x=0$ for all $|a|>\tau$. 
\end{lemma}
An example of a wave-function in $L^2(\mathbb{R})$ that satisfies all the conditions of the Lemma is $\phi(x)=1/\Gamma(\alpha+\beta x)\Gamma(\alpha-\beta x)$ for positive $\alpha$ and $\beta$, with $\alpha>1$. Using the results in \cite{napalkov}, it can be established that the complex extension $\phi(z)=1/\Gamma(\alpha+\beta z) \Gamma(\alpha-\beta z)$  is an entire function of exponential type $\pi\beta>0$.

Now let the probe wave-function $\braket{q}{\Psi_{Pr}}$ be momentum limited so that its complex extension is exponential of  type $\kappa_0$. Then the function $\braket{z-2\lambda b/\alpha}{\Psi_{Pr}}\braket{\Psi_{Pr}}{q-2\lambda b'/\alpha}$ is itself entire and exponential of type $2\kappa_0$ (see Methods). By the Lemma we have $I_{kl}(b,b')=0$ for all (finite) $\alpha>2\hbar\kappa_0/|a_k-a_l|$. Define $a_0=\mbox{min}\{(a_k-a_l), a_k>a_l\}$. Then the functions $I_{kl}(b,b')$ vanish for all $k\neq l$ when the system-probe coupling $\alpha$ satisfies 
\begin{equation}\label{decond}
\alpha>\alpha_D=\frac{2\hbar \kappa_0}{a_0} .
\end{equation}
Under this condition the reduced density matrix of the system and probe is diagonal and it assumes the form
\begin{equation}\label{reduced}
\rho_{S\otimes M}=\sum_{k} \left|\braket{\varphi_k}{\psi_0}\right|^2 \ketbra{\varphi_k}{\varphi_k}\otimes \rho_k ,
\end{equation}
where  $\rho_k$ is the state of the pointer after the measurement and is given by
\begin{equation}
\rho_k = \int \!\! \mbox{d}p \left|\braket{p}{\Psi_{Pr}}\right|^2 e^{\frac{i}{\hbar}(\alpha a_k -p)\beta B} \ketbra{\Phi_0}{\Phi_0} e^{\frac{i}{\hbar}(\alpha a_k -p)\beta B} .
\end{equation}
In \ref{reduced} perfect correlation has been established between eigenstates $\ket{\varphi_k}$ of $A$ and the set of states $\rho_k$ of the pointer. The states $\rho_k$'s then constitute the pointer states. 

We have thus achieved the statistical collapse of the density matrix at finite coupling strength; moreover, the collapse occurs right after the measurement, i.e. within a finite time interval. However, the pointer states are generally mixed and non-orthogonal. We now show that under similar conditions as above the $\rho_k$'s are pairwise orthogonal. For $k\neq l$ we have
\begin{eqnarray}
\rho_k \rho_l &=& \iint db \, db' e^{\frac{i}{\hbar} \lambda (a_k b - a_l b')} \braket{b}{\Phi_0} \braket{\Phi_0}{b'} \nonumber \\
&& \hspace{12mm} \times S_{k,l}(b,b') \ketbra{b}{b'}
\end{eqnarray}
where
\begin{eqnarray}
S_{k,l}(b,b')&=&\int_{-\infty}^{\infty} \mathrm{d}b'' e^{\frac{i}{\hbar} \lambda (a_l-a_k)b''} \abs{\braket{b''}{\Phi_0}}^2\nonumber\\
&&\hspace{10mm} \times  F(b'-b'') F(b''-b) .
\end{eqnarray}
$S_{k,l}(b,b')$ controls the orthogonality of $\rho_k$ and $\rho_l$; and  $\rho_k$ and $\rho_l$ are orthogonal if $S_{k,l}(b,b')=0$ for all $b,b'$. To avoid unnecessary complications, we can assume that $F(\eta)$ is real and even, i.e. $F(\eta)=F(-\eta)$. This requires that $\braket{p}{\Psi_{Pr}}$ is real and with definite parity. For $k\neq l$ equation is again a Fourier transform. 

By the same reasoning above, we can make $S_{k,j}(b,b')$ vanish by imposing that $\braket{b}{\Phi_0}$ is momentum limited of some  type $b_0$, which we assume to be real as well. Now $F(\eta)$ is exponential of type $2\lambda \kappa_0/\alpha$. By the Lemma the integral vanishes when $\lambda|a_l-a_k|>\hbar(2 b_0+ 4 \lambda \kappa_0/\alpha)$. Since $\lambda\kappa_0/\alpha$ is positive, the condition implies that the coupling constant $\lambda$ cannot be arbitrary but must satisfy the condition
\begin{equation}\label{pe}
\lambda>\lambda_0 = \frac{2\hbar b_0}{a_0}
\end{equation}
Then for all $k$ and $l$, $k\neq l$, orthogonality is achieved under the condition
\begin{equation}\label{orcond}
\alpha > \alpha_0 = \frac{4\hbar \kappa_0}{a_0 - 2\hbar b_0/\lambda}
\end{equation}
together with equation (\ref{pe}).

The orthogonality of the pointer states implies that we can associate then with an observable of the pointer. For fixed $k$ let $S_k$ be the support of $\rho_k$, i.e. $S_k$ comprises all vectors $\ket{\phi}$ of $\mathcal{H}_{P_0}$ such that $\rho_k\ket{\phi}\neq 0$. Also let $S_0$ be the simultaneous kernels of the $\rho_k$'s, i.e. $S_0$ comprises all vectors $\ket{\varphi}$ of $\mathcal{H}_{P_0}$ such that $\rho_k\ket{\varphi}=0$. $S_0$ may possibly constitute only the zero vector. Then the pointer Hilbert space has the direct sum decomposition $\mathcal{H}_{Po}=S_0\oplus S_1\oplus \dots \oplus S_N$, the subspaces $S_k$'s being mutually orthogonal. This allows us to associate a projector $\Pi_k$ to each $S_k$, $k=0,1,\dots,N$, where $\sum_k \Pi_k = \mathbb{I}_{Po}$. The set $\{\Pi_k\}$ is a projection valued measure, and it constitutes an observable of the pointer. The set of states $\rho_k$ then corresponds to a sharp observable.

Comparing equations (\ref{decond}) and (\ref{orcond}), we find that $\alpha_0>\alpha_D$  by virtue of equation (\ref{pe}). Hence once the orthogonality condition (\ref{orcond}) is satisfied the decoherence condition  is already satisfied. However, the condition on the induced coupling constant $\lambda$  given by equation (\ref{pe}) must first be satisfied before the decoherence and orthogonality conditions are met. Conditions (\ref{decond}), (\ref{pe}), and (\ref{orcond}) lend to the interpretation as requiring the minimum impulse that the probe has to transfer to the system and to the pointer in the course of the measurement, together with the minimum induced impulse between the system and the pointer. They show that it is only possible to have completely decohering and orthogonal measurement for finite measuring time and coupling strengths provided the probe and the pointer are prepared in momentum limited states. Thus we have shown the  existence of a measurement scheme that accomplishes the goal of environment induced decoherence theory  without the aid of an environment. 

{\bf Methods} 

{\it Measurement dynamics:} Let $H(t)$ be a time dependent Hamiltonian that commutes at different times, i.e.  $[H(t),H(t')]=0$  for all $t,t'$. Then the solution to the Schr\"{o}dinger equation is given by
\begin{equation}
\ket{\psi(t)}=\exp\left(-\frac{i}{\hbar}\int_{t_0}^{t}\mbox{d}\tau H(\tau)\right)\ket{\psi(t_0)}
\end{equation}
Quantum measurement is modeled by this kind of Hamiltonian, in particular, of the form $H(t)=g(t) H_M$, where $g(t)$  has compact support and $H_M$  is time independent; we will assume that $\int_{t_1}^{t_2}g(t)\mbox{d}t=1$, where the interval $[t_1,t_2]$  is the support of $g(t)$, so that $\Delta t=t_2-t_1$ is the duration of the measurement process. Then the state of the system at  right after the measurement is given by 
\begin{equation}
\ket{\psi(t_2)}=\mbox{e}^{-i H_M/\hbar} \ket{\psi(t_1)}
\end{equation}
 There are two situations at which this relation holds: First, is when the free Hamiltonian of the systems involved are zero; and second is when $\Delta t$ is much smaller than any relevant time scales of the free systems involved. The last one is true in particular when the measurement is impulsive. In the paper, we referred to $H_M$ as the measurement Hamiltonian even though it does not have the unit of energy.

{\it Proof of the Lemma}: For $a>0$ perform the contour integration $\oint \mbox{e}^{i a z} f(z)\mbox{d}z$ around the rectangle with vertices at  $[-L,0]$, $[L,0]$,  $[L,L+i\gamma]$, $[-L,-L+i \gamma]$; and then take the limit as $L\rightarrow\infty$. If $f(z)$ is exponential of type $\tau$  and if $\int_{-\infty}^{\infty} |f(x)| \mbox{d}x=M<\infty$, it is known that $\int_{-\infty}^{\infty} |f(x+i y)| \mbox{d}x\leq M \mbox{e}^{\tau |y|}$ \cite{boas}. Under this condition, the integral along the edges parallel to the complex axis will vanish in the said limit. Since $f(z)$  is entire, it has no pole so that we have the equality $\int_{-\infty}^{\infty} \mbox{e}^{i a x} f(x) \mbox{d}x= \mbox{e}^{-a\gamma}\int_{-\infty}^{\infty} \mbox{e}^{i a x} f(x+i \gamma) \mbox{d}x$  for all $\gamma>0$. From this follows the inequality  $|\int_{-\infty}^{\infty}\mbox{e}^{i a x} f(x) \mbox{d}x|\leq M \mbox{e}^{-\gamma(a-\tau)}$. The right hand side of the inequality can be made arbitrarily small by making $\gamma$  arbitrarily large. This implies that the integral vanishes. For $a<0$  close the contour in the lower half plane to obtain the same result.

{\it Product of two exponential type functions:} Let $f(z)$ and $g(z)$ be exponential functions of types $\tau_1$ and $\tau_2$, respectively, whose restrictions in the real line, $f(x)$ and $g(x)$, are vectors of $L^2(\mathbb{R})$. Then they can be written as $f(z)=\int_{-\tau_1}^{\tau_1} \mbox{e}^{i z t}\tilde{f}(t)\mbox{d}t$ and $g(z)=\int_{-\tau_2}^{\tau_2} \mbox{e}^{i z t} \tilde{g}(t)\mbox{d}t$, so that $f(z) g(z)=\int_{-(\tau_1+\tau_2)}^{(\tau_1+\tau_2)} \mbox{e}^{i z u} h(u) \mbox{d}u$, where $h(u)\propto\int_{-(\tau_1+\tau_2)}^{(\tau_1+\tau_2)} \tilde{f}((u+v)/2) \tilde{g}((u-v)/2) \mbox{d}v$. Hence $f(z)g(z)$ is exponential of type $(\tau_1+\tau_2)$.

\end{document}